\documentclass[11pt]{article}
\usepackage{acl}
\usepackage{times}
\usepackage{latexsym}
\usepackage[T1]{fontenc}
\usepackage[utf8]{inputenc}
\usepackage{microtype}
\usepackage{inconsolata}

\usepackage{graphicx}
\usepackage{booktabs}
\usepackage{amsmath}
\usepackage{amssymb}
\usepackage{multirow}
\usepackage{listings}
\usepackage{tikz}
\usetikzlibrary{shapes.geometric, shapes.symbols, arrows.meta, positioning, fit, backgrounds, calc}

\author{
Julius Näumann$^{\ast}$ \hspace{0.2cm}
Sven Keidel$^{\ast}$ \hspace{0.2cm}
Amir Molzam Sharifloo$^{\ast}$ \hspace{0.2cm}
Mira Mezini$^{\ast\dagger\ddagger}$ \\
\\
$^{\ast}$TU Darmstadt \\
$^{\dagger}$Hessian Center for Artificial Intelligence, Darmstadt, Germany \\
$^{\ddagger}$National Research Center for Applied Cybersecurity ATHENE
}

\title{Beyond BLEU: A Semantic Evaluation Method for Code Translation}
\begin{document}\maketitle

\begin{abstract}
Code translation is one of the core capabilities of LLMs. However, evaluating the correctness of translations remains difficult, as commonly used metrics such as BLEU measure only syntactic similarity, disregarding program semantics. We propose a novel evaluation methodology for code translation tasks, emphasizing semantic equivalence over surface-level string similarity. Our approach applies established compiler testing methodology to a new domain, allowing the assessment of an LLM fine-tuned for binary lifting tasks (i.e. decompiling binaries to higher-level representations). We introduce a semantic correctness score, defined as the proportion of translations that produce correct execution outcomes, and demonstrate its application by evaluating LLM-based and heuristic decompilers. Our findings show that LLM-based approaches significantly outperform heuristic ones, while BLEU scores show negligible correlation with semantic correctness ($r=-0.127$ to $0.354$), demonstrating that syntactic metrics fail to predict functional accuracy.
\end{abstract}

\section{Introduction}
Code translation is a key capability of large language models (LLMs), supporting a range of applications from cross-language interoperability \citep{dhruv2025} to automated decompilation \citep{hu2024}. Prior studies have often relied on \textit{string similarity metrics}, e.g., the BLEU score, to assess the correctness of code translations, measuring surface-level resemblance between the translated and original programs \citep{lei2023, aljagthami2025, cummins2025}. More sophisticated approaches such as RUBY \citep{tran2019} and CodeBLEU \citep{ren2020} consider code structure and incorporate AST in the calculation, but do not consider execution semantics.

Such syntactic metrics are ill-suited for semantic-preserving decompilation tasks, which require evaluating semantic equivalence, i.e., that translated programs faithfully replicate the original program's semantics, rather than syntactic similarity: (a) programs that are functionally identical may differ in syntax, choice of identifiers, or structural organization, yet string similarity assigns them a low score, thereby underestimating correctness; (b) programs that share superficial textual patterns can achieve high similarity scores even when their behaviors diverge, resulting in misleading overestimation.

Another approach is to assess the correctness of code translation with tests \citep{chen2021}. These tests run the translated program to check if it has the same functionality as the original program. This approach requires manual effort for creating each test case, and tests may not be exhaustive, leaving the semantic equivalence of untested code in question.

In this work, we present a scalable methodology for evaluating semantic-preserving code translation, addressing the limitations of prior methods. Our approach automatically generates random programs, eliminating manual effort and enabling the creation of arbitrarily large evaluation sets. The generated programs compute check-sums after each instruction, offering greater reliability than string-based similarity metrics in checking of semantic equivalence between original and translated code. 

We realize the new methodology using Csmith, a tool originally designed for automated compiler testing \citep{yang2011a}, which produces compilable C programs with integrated checksum logic. This enables systematic, reproducible, and fine-grained evaluation of semantic-preserving code translation tasks. We use our evaluation methodology to assess the correctness of LLM-based decompilers, specifically, Meta's LLM Compiler \citep{cummins2025}\footnote{Made available under \citeauthor{meta2025} \citeyear{meta2025}}, in the task of decompiling from x86 assembly to LLVM-bytecode. Our evaluation reveals that BLEU scores show no correlation with semantic correctness (point-biserial $r=-0.13$ to 0.35).

In summary, we contribute (1) an evaluation methodology for code translation based on automated program generation and semantic checksum validation, (2) a resulting semantic correctness score providing a direct measure of functional accuracy, and (3) empirical evidence that BLEU score is unrelated to semantic equivalence, showing the need for more sophisticated evaluation methods.

\section{Assessing Program Equivalence with String Similarity Metrics}
Decompilers translate the code of a low-level language to the code of a higher-level language, e.g., from x86 assembly to LLVM byte code \citep{avast2025, yadavalli2019}. An important property of decompilers is to preserve the semantics of the original program. More formally, two programs are observationally equivalent, if for all sequences of inputs they return the same sequence of outputs \citep{plotkin1977}. In general, deciding computationally if two programs have the same semantics is impossible due to the halting problem \citep{rice1953}.

As an approximate metric for semantic equivalence of programs, researchers have used string similarity metrics \citep{lei2023, aljagthami2025, evtikhiev2023, tran2019}. Specifically, the decompiled program is recompiled to the same language as the original program and then compared with a string similarity metric:

% --- The generic diagram (before Problem 1) ---
\begin{center}
\begin{tikzpicture}[
    node distance=0.6cm and 1.6cm,
    box/.style={draw, rectangle, align=left, font=\small\rmfamily, inner sep=5pt}
]
\node[box] (orig) {Original\\Program};
\node[box, below=of orig] (decomp) {Decompiled\\Program};
\node[box, right=of decomp] (recomp) {Recompiled\\Program};

\draw[->, dotted] (orig) -- (decomp);
\draw[->, dotted] (decomp) -- (recomp);

\coordinate (corner) at (recomp.north |- orig.east);
% Node placed on the horizontal segment
\draw[<-] (orig.east) -- node[above, font=\small\rmfamily] {String Similarity} (corner) -- (recomp.north);
\end{tikzpicture}
\end{center}

Two problems arise when using string-similarity as an approximation for program equivalence:

\textbf{Problem 1: Programs are semantically equivalent, but get low string similarity score.} Recompilation may optimize the program, leading to a low BLEU score between recompiled and original program:

% --- Problem 1 Diagram ---
\begin{center}
\begin{tikzpicture}[
    node distance=0.6cm and 0.8cm,
    box/.style={draw, rectangle, align=left, font=\small\ttfamily, inner sep=5pt}
]
\node[box] (orig) {mov eax, 5\\inc eax\\ret};
\node[box, below=of orig] (decomp) {\%1 = add i32 5, 1\\ret i32 \%1};
\node[box, right=of decomp] (recomp) {mov eax, 6\\ret};

\draw[->, dotted] (orig) -- (decomp);
\draw[->, dotted] (decomp) -- (recomp);

\coordinate (corner) at (recomp.north |- orig.east);
% Node placed on the horizontal segment
\draw[<-] (orig.east) -- node[above, font=\small\rmfamily] {BLEU-1: 0.54} (corner) -- (recomp.north);
\end{tikzpicture}
\end{center}

\vspace{0.3cm}
\textbf{Problem 2: Programs are not semantically equivalent, but receive high string similarity score.} A slight syntactic variation may lead to different semantics, despite high BLEU score. A decompiler may wrongly interpret the \texttt{lea} instruction as a load from memory, which does not have the same semantics.

% --- Problem 2 Diagram ---
\begin{center}
\resizebox{0.95\columnwidth}{!}{
\begin{tikzpicture}[
    node distance=0.6cm and 0.8cm,
    box/.style={draw, rectangle, align=left, font=\small\ttfamily, inner sep=5pt}
]
\node[box] (orig) {lea rax, [rdi+rsi]\\mov rbx, rax\\ret};
\node[box, below=of orig] (decomp) {\%1 = getelementptr i8, ptr\\\%rdi, i64 \%rsi\\\%2 = load i64, ptr \%1\\ret i64 \%2};
\node[box, right=of decomp] (recomp) {mov rax, [rdi+rsi]\\mov rbx, rax\\ret};

\draw[->, dotted] (orig) -- (decomp);
\draw[->, dotted] (decomp) -- (recomp);

\coordinate (corner) at (recomp.north |- orig.east);
% Node placed on the horizontal segment
\draw[<-] (orig.east) -- node[above, font=\small\rmfamily] {BLEU-1: 0.92} (corner) -- (recomp.north);
\end{tikzpicture}
}
\end{center}

These examples show that string similarity metrics are unsuitable to assess if programs are semantically equivalent.

\section{Methodology}
In the field of compiler testing, semantic equivalence is a rigorous requirement used to validate the correctness of optimization steps: a program should compile to a binary with identical behavior regardless of the applied optimizations. A common way to test this property is through the automatic generation of programs whose executions can be compared to detect discrepancies. If two binaries produced from the same source yield different outcomes, the transformation has altered the program's semantics. Our evaluation methodology is inspired by this approach and implemented for the C language, though it is generalizable to any language.

We use Csmith \citep{yang2011b} to generate random C programs performing arbitrary computations and function calls. Binaries are produced at optimization levels -O0 and -O3. Each program produces a CRC checksum over intermediate values when compiled and executed; differing checksums indicate non-equivalent semantics. This mechanism allows us to evaluate the translation accuracy of binary lifters.

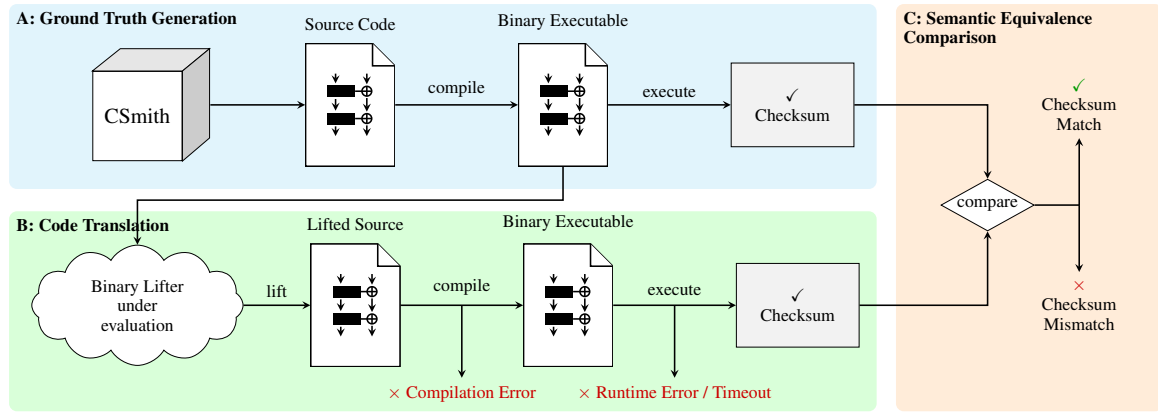
\begin{figure*}[tb]
\centering
\resizebox{2\columnwidth}{!}{
\begin{tikzpicture}[
    node distance=1.5cm and 1.5cm,
    box/.style={draw, rectangle, minimum width=2.2cm, minimum height=1.5cm, align=center, fill=gray!10, font=\small},
    cloud_node/.style={draw, cloud, cloud puffs=12, cloud puff arc=120, aspect=2.2, minimum width=3cm, minimum height=1.6cm, align=center, fill=white, font=\small},
    % Custom shape for documents with a folded top-right corner
    doc/.style={
        minimum width=1.6cm, minimum height=2.2cm, align=center,
        path picture={
            \filldraw[fill=white, draw=black, thick]
                (path picture bounding box.south west) --
                (path picture bounding box.north west) --
                ($(path picture bounding box.north east) + (-0.4, 0)$) --
                ($(path picture bounding box.north east) + (0, -0.4)$) --
                (path picture bounding box.south east) -- cycle;
            \filldraw[fill=white, draw=black, thick]
                ($(path picture bounding box.north east) + (-0.4, 0)$) --
                ($(path picture bounding box.north east) + (-0.4, -0.4)$) --
                ($(path picture bounding box.north east) + (0, -0.4)$);
        }
    },
    arrow/.style={-stealth, thick},
    decision/.style={draw, diamond, aspect=1.8, align=center, fill=white, font=\small, inner sep=1pt}
]

% Perfectly centered internal graphic for the document files
\tikzset{
    filecontent/.pic={
        \fill[black] (-1.15, 0.4) rectangle (0.25, 1.0);
        \fill[black] (-1.15, -1.0) rectangle (0.25, -0.4);
        \draw[-stealth, thick] (-0.75, 1.6) -- (-0.75, 1.1);
        \draw[-stealth, thick] (-0.75, 0.3) -- (-0.75, -0.3);
        \draw[-stealth, thick] (-0.75, -1.1) -- (-0.75, -1.6);
        \draw[-stealth, thick] (0.85, 1.6) -- (0.85, 1.05);
        \draw[-stealth, thick] (0.85, 0.35) -- (0.85, -0.35);
        \draw[-stealth, thick] (0.85, -1.05) -- (0.85, -1.6);
        \draw[thick] (0.85, 0.7) circle (0.25);
        \draw[thick] (0.85, 0.45) -- (0.85, 0.95);
        \draw[thick] (0.6, 0.7) -- (1.1, 0.7);
        \draw[thick] (0.85, -0.7) circle (0.25);
        \draw[thick] (0.85, -0.95) -- (0.85, -0.45);
        \draw[thick] (0.6, -0.7) -- (1.1, -0.7);
        \draw[thick] (0.25, 0.7) -- (0.6, 0.7);
        \draw[thick] (0.25, -0.7) -- (0.6, -0.7);
    }
}

% --- ZONE A: Ground Truth ---
% 3D Cube for CSmith with darker shaded top and right faces
\node[draw, fill=white, minimum width=1.6cm, minimum height=1.6cm] (csmith) {CSmith};
\draw[fill=black!10] (csmith.north west) -- ++(0.5,0.4) coordinate (nw) -- ++(1.6,0) coordinate (ne) -- (csmith.north east) -- cycle;
\draw[fill=black!20] (csmith.north east) -- (ne) -- ++(0,-1.6) coordinate (se) -- (csmith.south east) -- cycle;

% Documents and Process A (Shifted slightly up to align with 3D cube center)
\node[doc, right=2.2cm of csmith, yshift=0.2cm] (src1) {};
\pic at (src1.center) [scale=0.35] {filecontent};
\node[above=0.1cm of src1, font=\small] (l_src1) {Source Code};

\node[doc, right=2.2cm of src1] (bin1) {};
\pic at (bin1.center) [scale=0.35] {filecontent};
\node[above=0.1cm of bin1, font=\small] (l_bin1) {Binary Executable};

\node[box, right=2.2cm of bin1] (chk1) {\checkmark \\ Checksum};

% Arrow originates from the exact midpoint of the 3D cube's rightmost edge
\draw[arrow] ($(csmith.east) + (0.5, 0.2)$) -- (src1);
\draw[arrow] (src1) -- node[above, font=\small] {compile} (bin1);
\draw[arrow] (bin1) -- node[above, font=\small] {execute} (chk1);

% --- ZONE B: Code Translation ---
\node[cloud_node, below=1.5cm of csmith] (lifter) {Binary Lifter \\ under \\ evaluation};

\node[doc, right=1.2cm of lifter] (src2) {};
\pic at (src2.center) [scale=0.35] {filecontent};
\node[above=0.1cm of src2, font=\small] (l_src2) {Lifted Source};

\node[doc, right=2.2cm of src2] (bin2) {};
\pic at (bin2.center) [scale=0.35] {filecontent};
\node[above=0.1cm of bin2, font=\small] (l_bin2) {Binary Executable};

\node[box, right=2.2cm of bin2] (chk2) {\checkmark \\ Checksum};

% Link from A down to B (Routes perfectly in the negative space between the boxes)
\draw[arrow] (bin1.south) -- ++(0, -0.6) -| (lifter.north);

% Process B Arrows
\draw[arrow] (lifter) -- node[above, font=\small] {lift} (src2);

% Compilation and Execution with error branches
\draw[arrow] (src2) -- coordinate (comp_mid) (bin2);
\node[above=0.05cm of comp_mid, font=\small] {compile};
\draw[arrow] (comp_mid) -- ++(0, -1.3) node[below, font=\small, text=red!80!black] {$\times$ Compilation Error};

\draw[arrow] (bin2) -- coordinate (exec_mid) (chk2);
\node[above=0.05cm of exec_mid, font=\small] {execute};
\draw[arrow] (exec_mid) -- ++(0, -1.3) node[below, font=\small, text=red!80!black] {$\times$ Runtime Error / Timeout};

% --- ZONE C: Semantic Equivalence ---
\path (chk1.east) -- coordinate (chk_mid) (chk2.east);
\node[decision, right=1.5cm of chk_mid] (comp) {compare};

% Arrows routing horizontally then dropping into the top/bottom points
\draw[arrow] (chk1.east) -| (comp.north);
\draw[arrow] (chk2.east) -| (comp.south);

% Lines out of diamond
\draw[thick] (comp.east) -- ++(0.8,0) coordinate (split);
\draw[arrow] (split) -- ++(0, 1.2) node[above, align=center, font=\small] {\textcolor{green!60!black}{\checkmark} \\ Checksum \\ Match};
\draw[arrow] (split) -- ++(0, -1.2) node[below, align=center, font=\small] {\textcolor{red!80!black}{$\times$} \\ Checksum \\ Mismatch};

% --- BACKGROUND LAYERS (Grid Aligned) ---
\begin{scope}[on background layer]
    % Zone A (Blue) - Stretched left to match the cloud boundary
    \coordinate (A_nw) at ($(lifter.west |- nw) + (-0.4, 0.8)$);
    \coordinate (A_se) at ($(chk1.south east |- bin1.south) + (0.4, -0.4)$);
    \fill[cyan!10, rounded corners] (A_nw) rectangle (A_se);
    \node[anchor=north west, font=\small\bfseries] at (A_nw) {A: Ground Truth Generation};

    % Zone B (Green) - Tucked cleanly below the routing arrow
    \coordinate (B_nw) at ($(A_nw |- lifter.north) + (0, 0.6)$); 
    \coordinate (B_se) at ($(A_se |- comp_mid) + (0, -1.9)$);
    \fill[green!15, rounded corners] (B_nw) rectangle (B_se);
    \node[anchor=north west, font=\small\bfseries] at (B_nw) {B: Code Translation};

    % Zone C (Orange) - Perfect Y-axis grid alignment with A and B
    \coordinate (C_nw) at ($(comp.west) + (-0.8, 0)$);
    \coordinate (C_nw_aligned) at (C_nw |- A_nw);
    \coordinate (C_se) at ($(split) + (1.5, 0)$);
    \coordinate (C_se_aligned) at (C_se |- B_se);
    \fill[orange!15, rounded corners] (C_nw_aligned) rectangle (C_se_aligned);
    \node[anchor=north west, font=\small\bfseries, align=left] at (C_nw_aligned) {C: Semantic Equivalence\\Comparison};
\end{scope}

\end{tikzpicture}
}
\caption{An overview of the evaluation framework.}
\label{fig:framework}
\end{figure*}

Figure \ref{fig:framework} illustrates the process. Csmith generates input programs, which are compiled to obtain reference checksums. The binaries are then processed by the decompiler under evaluation to produce source code, which is compiled and executed. The resulting checksum is compared to the reference to determine whether the lifted binary is semantically equivalent to the original. The approach identifies and reports potential points of failure throughout the process, from decompiling to compilation and execution.

\textbf{Semantic Correctness Score.} We define the semantic correctness score as the proportion of successfully translated programs that produce matching checksums when executed:

\begin{equation}
\text{Semantic Score} = \frac{\text{\# programs w/ correct checksum}}{\text{\# tested programs}}
\end{equation}
A score of 1.0 indicates perfect semantic preservation, while lower scores reflect the frequency of semantic errors.

\section{Evaluating the correctness of Decompilers}

\textbf{Study Setup.} We apply our methodology to evaluate the accuracy of Meta's LLMCompiler model and two widely used state of the art heuristic lifters, RetDec \citep{avast2025} and McToll \citep{yadavalli2019}. In total, we generated 1024 test files, constrained to fit the model's 8192 token context window and excluding trivial programs. Csmith generation and program execution are deterministic, CPU-based processes requiring only seconds per program, producing only minimal overhead. We evaluate the 13b parameter LLMCompiler model, fine-tuned on the binary lifting task, at a temperature of 1.0. We perform a single experiment run, on two H100 GPUs, using a total of 88 GPU hours.

\textbf{Comparative Performance.} LLMCompiler substantially outperforms heuristic-based decompilers, achieving semantic correctness scores of 0.33 and 0.63 on O0 and O3 binaries, respectively. Neither Mctoll nor Retdec were able to produce a binary with successful execution, with Mctoll failing to lift any binaries, and RetDec producing only binaries that either failed to compile or crashed at runtime.

\textbf{BLEU Scores Do Not Predict Semantic Correctness.} We compute round-trip BLEU scores for all programs for which LLMCompiler produced compilable outputs. Table \ref{tab:bleu_corr} lists point-biserial correlation coefficients between these similarity scores and semantic correctness, where values near 1 or -1 indicate strong predictive power and 0 indicating no correlation. 

\begin{table}[h!]
\centering
\caption{LLMCompiler Round-Trip Similarity vs. Execution Result.}
\label{tab:bleu_corr}
\resizebox{\columnwidth}{!}{
\begin{tabular}{ccccc}
\toprule
\textbf{\begin{tabular}[c]{@{}c@{}}Opt. Level\\ Pass/Fail\end{tabular}} & \textbf{Metric} & \textbf{\begin{tabular}[c]{@{}c@{}}Pass\\ Mean\end{tabular}} & \textbf{\begin{tabular}[c]{@{}c@{}}Fail\\ Mean\end{tabular}} & \textbf{Corr. (r)} \\
\midrule
\multirow{3}{*}{\begin{tabular}[c]{@{}c@{}}O0\\ 338/228\end{tabular}} & BLEU-1 & 0.81 & 0.84 & $-0.13^{**}$ \\
 & BLEU-4 & 0.77 & 0.77 & $-0.03$ \\
 & CodeBLEU & 0.84 & 0.85 & $-0.05$ \\
\midrule
\multirow{3}{*}{\begin{tabular}[c]{@{}c@{}}O3\\ 643/276\end{tabular}} & BLEU-1 & 0.71 & 0.71 & $0.02$ \\
 & BLEU-4 & 0.60 & 0.57 & $0.18^{***}$ \\
 & CodeBLEU & 0.80 & 0.75 & $0.35^{***}$ \\
\bottomrule
\multicolumn{5}{l}{Significance: $^* p < 0.05$, $^{**} p < 0.01$, $^{***} p < 0.001$} \\
\end{tabular}
}
\end{table}

The results reveal that BLEU scores fail to predict semantic correctness, with the strongest predictor (CodeBLEU at O3) showing only weak correlation. Mean similarity scores for succeeding and failing testcases are very close, indicating a near-overlap in distributions.

\section{Risks and Limitations}
Given that our framework builds on Csmith, it is currently limited to assessing semantic equivalence between only programs generated from C code and those derived from it. Including support for other languages would entail a rewrite of a comparable tool, accounting for any language-specific constructs and semantics. Moreover, the code generated by our framework may lie outside of the distribution of training data used to train LLMCompiler. Nonetheless, the framework still provides a valuable measure of performance. A high semantic score would indicate strong applicability and a high level of semantic equivalence, and importantly, a good model should not be limited to its training data.

\bibliography{references}

\clearpage
\onecolumn
\appendix
\section{Appendix}
This appendix provides detailed evaluation results supporting our findings that (1) LLM-based code translation substantially outperforms heuristic approaches when measured by semantic correctness, and (2) syntactic similarity metrics fail to predict functional accuracy.

\subsection{Complete Evaluation Results}
Table \ref{tab:full_eval} presents a comprehensive breakdown of all 1024 test programs across three lifters (LLMCompiler, McToll, RetDec) at two optimization levels (O0, O3). The table categorizes outcomes into distinct failure modes:
\begin{itemize}
    \item \textbf{Lifting error:} The lifter failed to produce any output code
    \item \textbf{Compilation error:} Output code contained syntax errors preventing compilation
    \item \textbf{Runtime error:} Compiled code crashed (segmentation fault) or timed out (infinite loop)
    \item \textbf{Checksum error:} Code executed successfully but produced incorrect output
    \item \textbf{Checksum correct:} Code executed and produced semantically equivalent output
\end{itemize}

\begin{table*}[h!]
\centering
\small
\caption{Summary of evaluation, comparing the effectivity of LLMCompiler and the heuristic lifters.}
\label{tab:full_eval}
\begin{tabular}{lcccccc}
\toprule
\textbf{Lifter} & \multicolumn{2}{c}{\textbf{LLMCompiler}} & \multicolumn{2}{c}{\textbf{mctoll}} & \multicolumn{2}{c}{\textbf{retdec}} \\
\textbf{Opt. Level} & \textbf{O0} & \textbf{O3} & \textbf{O0} & \textbf{O3} & \textbf{O0} & \textbf{O3} \\
\midrule
Tested programs & 1024 & 1024 & 1024 & 1024 & 1024 & 1024 \\
Lifting error & 0 & 0 & 1024 & 1024 & 0 & 0 \\
Compilation error & 105 & 458 & 0 & 0 & 181 & 38 \\
Compilation success & 566 (55.27\%) & 919 (89.75\%) & 0 (0.00\%) & 0 (0.00\%) & 843 (82.32\%) & 986 (96.29\%) \\
Runtime error & 26 & 12 & 0 & 0 & 843 & 982 \\
Checksum error & 264 & 202 & 0 & 0 & 0 & 4 \\
\textbf{Checksum correct} & \textbf{338 (33.01\%)} & \textbf{643 (62.79\%)} & \textbf{0 (0.00\%)} & \textbf{0 (0.00\%)} & \textbf{0 (0.00\%)} & \textbf{0 (0.00\%)} \\
\bottomrule
\end{tabular}
\end{table*}

\subsection{Visual Analysis of BLEU Score Distributions}
Figure 2 visualizes the relationship between round-trip BLEU scores and semantic correctness for LLMCompiler outputs. Each box plot shows the distribution of similarity scores for programs with correct checksums (green) versus incorrect checksums (red), across three metrics (BLEU-1, BLEU-4, CodeBLEU) and two optimization levels (O0, O3). The substantial overlap between green and red distributions demonstrates that BLEU scores cannot reliably distinguish semantically correct from incorrect translations. Programs with high BLEU scores (>0.9) frequently fail semantic validation, while programs with lower scores (<0.7) often succeed. This visual evidence corroborates the correlation analysis in Table \ref{tab:bleu_corr}, confirming that syntactic similarity is an unreliable proxy for functional correctness in code translation tasks.

\begin{figure*}[t]
\centering
% When the figure script is ready, import the pdf instead
% \includegraphics[width=\textwidth]{bleu_boxplots.pdf}
\includegraphics[width=\textwidth]{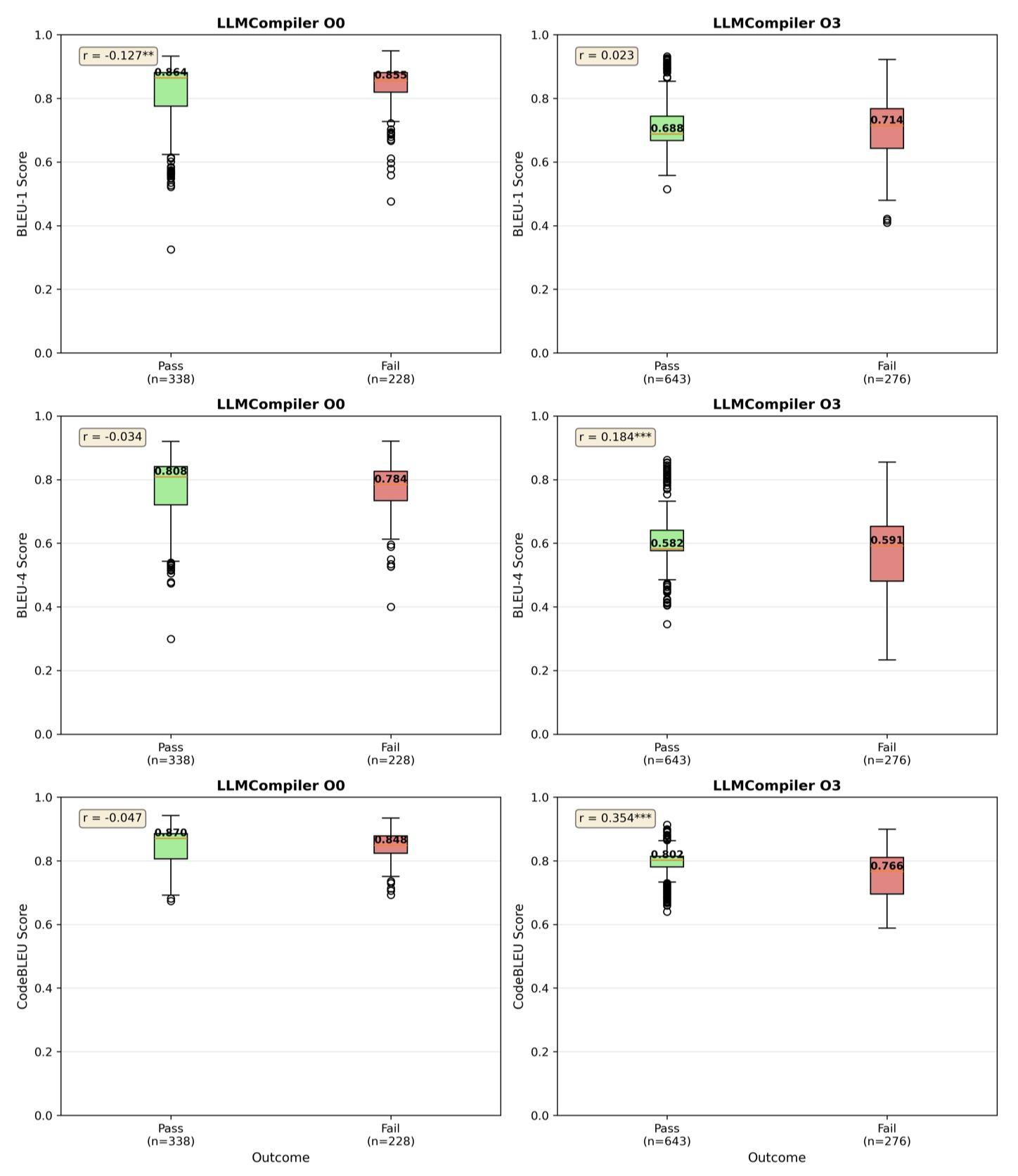}
\caption{BLEU similarity scores comparing original assembly to round-trip assembly (compiled from lifted source), with green and red plots representing a checksum match or mismatch respectively. The plots show a distribution overlap between checksum match and checksum fail string similarity scores.}
\label{fig:bleu_boxplots}
\end{figure*}

\end{document}